\documentclass[sigconf]{acmart}

\usepackage{pifont}
\usepackage{multirow}
\usepackage{tabularx}
\usepackage{ragged2e}
\AtBeginDocument{%
  \providecommand\BibTeX{{%
    \normalfont B\kern-0.5em{\scshape i\kern-0.25em b}\kern-0.8em\TeX}}}

\copyrightyear{2023}
\acmYear{2023}
\setcopyright{acmlicensed}\acmConference[MM '23]{Proceedings of the 31st ACM International Conference on Multimedia}{October 29-November 3, 2023}{Ottawa, ON, Canada}
\acmBooktitle{Proceedings of the 31st ACM International Conference on Multimedia (MM '23), October 29-November 3, 2023, Ottawa, ON, Canada}
\acmPrice{15.00}
\acmDOI{10.1145/3581783.3612046}
\acmISBN{979-8-4007-0108-5/23/10}

\begin{document}

\title{Dance with You: The Diversity Controllable Dancer Generation via Diffusion Models.}

\author{Siyue Yao}
\authornote{Equal contribution.}
\affiliation{%
  \institution{The Chinese University of Hong Kong, Shenzhen}
    \country{}
}
\email{yaosiyue1@gmail.com}

\author{Mingjie Sun}
\authornotemark[1]
\affiliation{%
  \institution{Soochow University}
  \country{}
}
\email{mjsun@suda.edu.cn}

\author{Bingliang Li}
\affiliation{%
  \institution{The Chinese University of Hong Kong, Shenzhen}
  \country{}
}
\email{bingliangli@link.cuhk.edu.cn}

\author{Fengyu Yang}
\affiliation{%
  \institution{The Chinese University of Hong Kong, Shenzhen}
  \country{}
}
\email{fengyuyang1@link.cuhk.edu.cn}

\author{Junle Wang}
\affiliation{%
  \institution{Tencent}
  \country{}
}
\email{wangjunle@gmail.com}

\author{Ruimao Zhang}
\authornote{Corresponding author.}
\affiliation{%
  \institution{The Chinese University of Hong Kong, Shenzhen}
  \country{}
}
\email{zhangruimao@cuhk.edu.cn}

\renewcommand{\shortauthors}{Siyue Yao et al.}

\begin{abstract}

Recently, digital humans for interpersonal interaction in virtual environments have gained significant attention. In this paper, we introduce a novel multi-dancer synthesis task called partner dancer generation, which involves synthesizing virtual human dancers capable of performing dance with users. The task aims to control the pose diversity between the lead dancer and the partner dancer. The core of this task is to ensure the controllable diversity of the generated partner dancer while maintaining temporal coordination with the lead dancer. This scenario varies from earlier research in generating dance motions driven by music, as our emphasis is on automatically designing partner dancer postures according to pre-defined diversity, the pose of lead dancer, as well as the accompanying tunes. To achieve this objective, we propose a three-stage framework called \textbf{Dan}ce-with-\textbf{Y}ou (\textbf{DanY}). Initially, we employ a 3D Pose Collection stage to collect a wide range of basic dance poses as references for motion generation. Then, we introduce a hyper-parameter that coordinates the similarity between dancers by masking poses to prevent the generation of sequences that are over-diverse or consistent. To avoid the rigidity of movements, we design a Dance Pre-generated stage to pre-generate these masked poses instead of filling them with zeros. After that, a Dance Motion Transfer stage is adopted with leader sequences and music, in which a multi-conditional sampling formula is rewritten to transfer the pre-generated poses into a sequence with a partner style. In practice, to address the lack of multi-person datasets, we introduce AIST-M, a new dataset for partner dancer generation, which is publicly availiable at \href{https://github.com/JJessicaYao/AIST-M-Dataset/}{https://github.com/JJessicaYao/AIST-M-Dataset}. Comprehensive evaluations on our AIST-M dataset demonstrate that the proposed DanY can synthesize satisfactory partner dancer results with controllable diversity.

\end{abstract}





\begin{teaserfigure}
\centering
\setlength{\abovecaptionskip}{0.2cm}
  \includegraphics[width=0.91\linewidth]{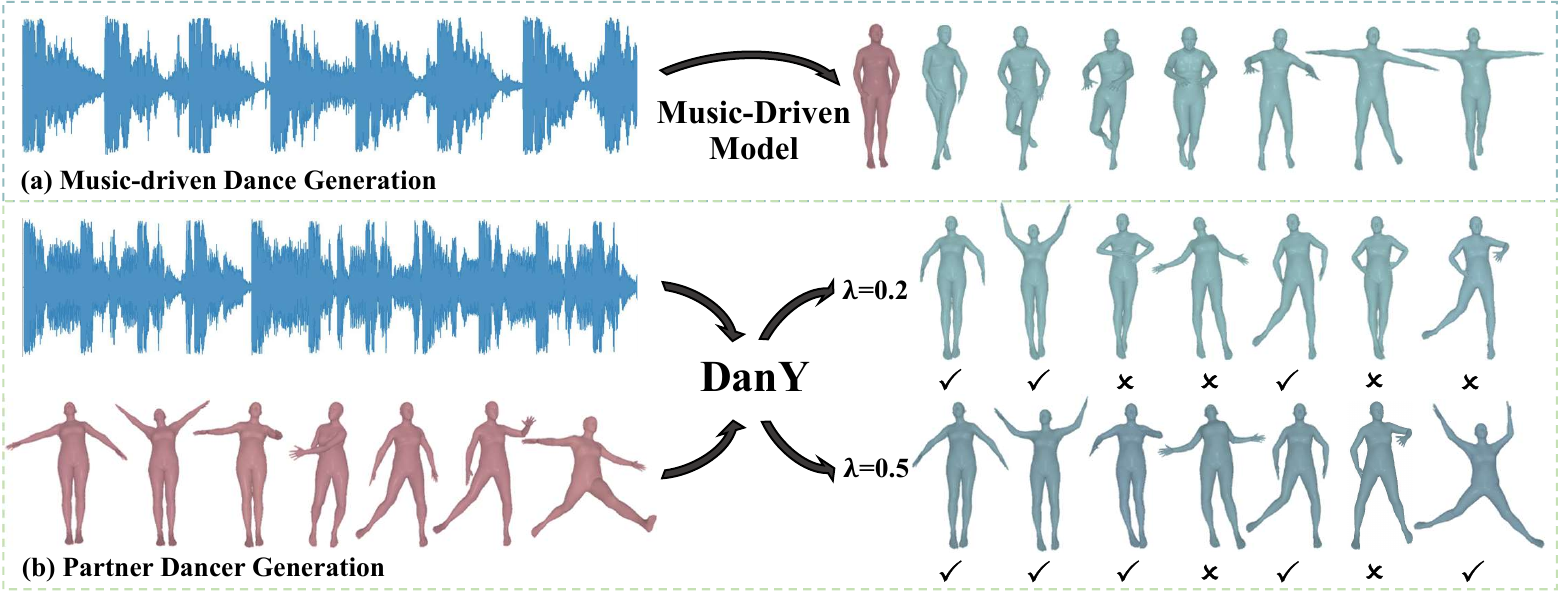}
  \caption{The comparison of our partner dancer generation task with previous music-driven dance generation task. (a) demonstrates the objective of the previous task, which is to generate a dance sequence from a piece of music and a start pose (the pink person). (b) is the objective of our partner dancer generation, which use a lead dancer sequence (the pink person) and a piece of music to create a partner dancer with diversity control. The green person and blue person in (b) are generated with similarity parameters $\lambda$ 0.2 and 0.5 given the same lead dancer and music, respectively. The check and cross symbols indicate whether the frame needs to be aligned with the lead dancer.}
  \label{define}
\end{teaserfigure}


\maketitle

\section{Introduction}
Dance, the earliest art produced by humans, has played a vital role in ritual, celebration and entertainment. With the emergence of the Metaverse, the creation of virtual dancers has become a significant trend on social media platforms. Many studies have been conducted to investigate how to create natural dancing movements, which are broadly employed in games, animation, film special effects, virtual idols and many other life scenes. 

Although dance generation methods have achieved various outcomes, the main focus is on creating a dance with a specific style according to the beat of music, as shown in Figure \ref{define}-(a). Most works \cite{ye2020choreonet, chen2021choreomaster,lee2019dancing, huang2020dance, zhuang2022music2dance, li2021ai, li2022danceformer} extract music feature as the guide to generate the single dancer frame by frame, given the first frame of the action. However, this strategy is not suitable for scenarios that require interactive creation with a partner dancer. For example, games that have additional game characters, like \textit{JUST DANCE} \cite{charbonneau2011teach}, are required to dance with players. Besides, virtual idols have been created to cooperate with other dancers and complete dance performances. Thus, the demand for deep human-computer interaction and collaboration in dance generation is increasing.

To address collaborative tasks in dance, we introduce a novel task named \textbf{Diversity Controllable Partner Dancer Generation} to create a virtual human dancing with a leader dancer. The task aims to address the complicated correlations between the lead and partner dancer, ensuring controllable diversity in the generated partner dancer and maintaining temporal coordination with the lead dancer. As shown in Figure \ref{define}-(b), the pink person is the lead dancer, while the green and blue person represent the generated dancers when the similarity parameter $lambda$ is 0.2 and 0.5. Our task is to utilize user-specific similarity to generate different partner dancers with the same lead dancer and music. Compared to previous works, our method focuses on achieving temporal consistency or difference between the movements of the lead and partner dancers. The main challenges of the task are (a) \textbf{the continuity of generation}: The generated motions need to be consistent with the lead dancer sequence at a certain time while maintaining continuity between motions without lagging; (b) \textbf{the controllable diversity}: The generated motions need to be coordinated with the lead dancer while ensuring the diversity of movements in the remaining moments; (c) \textbf{the datasets shortage}: The current available datasets are limited to single dancer and lack datasets for group dancers or lead-partner dancer pairs. 

To achieve the above objectives, a three-stage framework named \textbf{Dan}ce-with-\textbf{Y}ou (\textbf{DanY}) is introduced to realize the partner dance generation with expected diversity and consistency. The DanY contains a 3D Pose Collection stage, a Dance Pre-generated stage, and a Dance Motion Transfer stage. In the 3D Pose Collection stage, various dance poses are encoded to codebooks and used as references for motion generation. The Dance Pre-generated stage infers the consistent moments with a given similarity hyper-parameter and pre-generates the dance sequence in the remaining moments. In the last stage, the pre-generated dance movements are converted into sequences that are both consistent with the lead dancer movements for some parts and similar to the original partner dancer for the remaining parts by using the lead dancer motions and the input music as guidance. A multi-conditional sampling formula is rewritten to effectively control the depth architecture. 

In practice, we construct a 3D multiple dancer dataset named AIST-M for our partner dancer generation task, which is an extension of AIST \cite{tsuchida2019aist}. Our dataset annotates well-annotated 2D, 3D and Skinned Multi-Person Linear model (SMPL) \cite{loper2015smpl} format skeleton data. More importantly, we annotate the lead-partner dancer pair which has not been implemented in previous datasets. We also splice the lead-partner dancer pairs in our dataset according to the diversity control parameters as the ground truth of training, which is shown in detail in Section \ref{sec3.3}. Using the proposed dataset, we implement our DanY framework to generate partner dancer with controllable diversity. As shown in Figure \ref{define}-(b), with a difference ratio of similarity, the generated motion sequence exhibits more diversity with 0.2 (the green sequence) and more consistency with 0.5 (the blue sequence). The generation results demonstrate that our framework can provide an automated process for generating partner dancers in Metaverse or other virtual scenes, as well as provide valuable material for future multi-person choreography. 

In conclusion, we summarize our contributions as follows:
\begin{itemize}
\item We introduce a novel task named partner dancer generation, which aims to create virtual dancers that can perform dance with given lead dancer and music, enabling diversity control between the lead and partner dancer poses.
\item We organize a new AIST-M dataset based on group dance videos in the AIST dataset, which includes the lead dancer and partner dancer pairs with their 2D, 3D and SMPL format annotations. For now, AIST-M first distinguishes the roles between different dancers. Besides, we propose a new evaluation metric, named MFID, to evaluate the similarity of generation results.
\item  We propose a three-stage framework named DanY that allows for end-to-end diversity control using a rewritten diffusion sampling formula to construct partner dancer motions with music, lead dancer motions and a given controllable parameter. 
\end{itemize}

\section{Related Work}
\subsection{Dance Motion Synthesis}
Dance motion synthesis involves generating realistic and creative human motions, often guided by audio input. The traditional methods \cite{arikan2002interactive,kim2003rhythmic,lee2013music, fan2011example} use the music-motion constraints to ensure consistency by cropping and copying pieces of motion. Based on this pipeline, ChoreoMaster \cite{chen2021choreomaster} and ChoreoNet \cite{ye2020choreonet} map the relationship between rhythm signatures and dance pieces by deep learning. However, these methods lead to unnatural transitions and movement rigidity. To address this, researchers have proposed frameworks such as Chor-RNN \cite{crnkovic2016generative}, LSTM-based framework \cite{tang2018dance, huang2020dance}, MM-GAN \cite{lee2019dancing}, FACT \cite{li2021ai}, and Actor-Critic Motion GPT \cite{siyao2022bailando} to generate prolonged and realistic dance sequences via given musics.
However, these methods lack the ability to produce dances with multiple dancers. Motivated by this, GroupDancer \cite{wang2022groupdancer} and GDanceR \cite{le2023music} are proposed to tackle this challenge by  generating multiplayer dancers. Despite these advancements, there are still limitations in solving correlation and choreography between dancers in previous works.

\subsection{Diffusion Generative Models}
Diffusion model \cite{sohl2015deep, song2020improved} is an emerging generative model. Normally, it contains two parts: a forward process and a reverse process. The forward process gradually adds noise to samples, while a neural network is used to denoise in the reverse process. Ho \textit{et al.} \cite{ho2020denoising} apply the diffusion model in image generation and propose DDPM for sampling. Song \textit{et al.} \cite{song2020denoising} provide another way named DDIM in denoising sampling. 
With the requirements of conditional generation, classifier-guided diffusion \cite{dhariwal2021diffusion} is utilized which is guided by an explicit classifier.
Furthermore, the classifier-free guidance \cite{ho2021classifier} is proposed to avoid training an extra classification network. 
More recently, some researchers \cite{barquero2022belfusion,ahn2023can,chen2023humanmac} suggest predicting human motions by a diffusion framework. Besides, the methods \cite{zhang2022motiondiffuse, kim2022flame, tevet2022human} use diffusion models to generate or edit motion based on the given text. Motivated by this, we propose DanY for dancer generation based on the diffusion model.

\section{Methodology}
Given a 3D motion sequence of lead dancer $S_{l}= \{s_{l}^n \}^{N}_{n=1}$ with $N$ indicates the number of the frames, an accompanying music audio $M=\{m^n\}^{N}_{n=1}$ which means the music feature at the $n$ moment and a pre-defined similarity parameter $\lambda \in [0,1]$ with larger values implying lower diversity, our goal is to generate the partner dancer sequences $S_{p}=  \{s_{p}^n\}^{N}_{n=1} $ where the generated motion coincides with the timing of the lead dancer motion. Specifically, we represent any human pose $s \in \mathbf{R}^{J\times 3}$ as a $J\times 3$ dimensional vector where $J$ is the number of joints in the SMPL \cite{loper2015smpl} model and $3$ indicates 3D coordinates. 

\begin{figure*}[!t]
  \centering
  \setlength{\abovecaptionskip}{0.2cm}
  \includegraphics[width=0.9\linewidth, height=180pt]{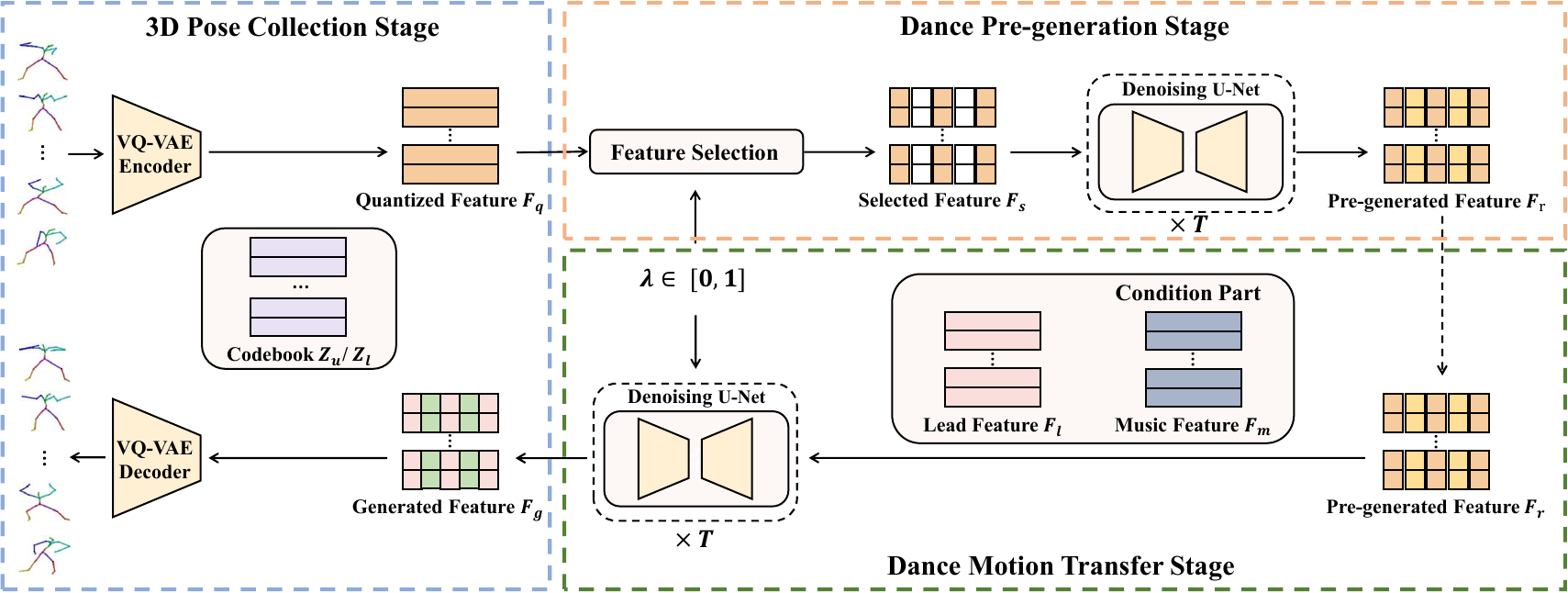}
  \caption{Partner Dancer Generation pipeline of DanY. In the 3D Pose Collection stage, a VQ-VAE encoder embeds the input sequence to the pose code. In the Dance Pre-generation stage, we introduce a similarity parameter $\lambda$ to control the aligned timestamps, and a diffusion model to pre-generate the pose. Afterwards, the feature with the characteristics of a partner dancer is generated via a Dance Motion Transfer stage according to the music features and lead dancer features. Finally, a VQ-VAE decoder is used to recover the generated feature to the partner sequence.}
  \label{overall}
\end{figure*}

To generate a partner dancer with controllable diversity, we proposed a three-stage framework DanY to synthesize the partner dancer as shown in Figure \ref{overall}. We define each stage as follows:
\begin{itemize}
\item \textbf{3D Pose Collection Stage.} Given sequence $S \in \mathbf{R}^{N \times J \times 3}$, an encoder is learned to quantize the sequence $S$ into the finite codebook $Z \in \mathbf{R}^{K \times C}$ where $K$ is the number of code and $C$ the channel dimension of features. A decoder is simultaneously trained to recover $\hat{S}$ from the quantized latent $F_q$.
This stage is used to collect the basic dance pose as well as to avoid variance due to the different speeds of the same pieces of motion.
\item \textbf{Dance Pre-generated Stage.} Having obtained the quantized latent $F_q$, a feature selection module is used to obtain the selected feature $F_s$, where $\lambda  \in [0,1] $ controls the number pose of inputs $F_q$ that should be selected. Then, the pre-generated latent $F_r$ is get from populating the selected inputs $F_s$. This stage is used to ensure the continuity of the final generated dance motions.
\item \textbf{Dance Motion Transfer Stage.} Given the pre-generated latent $F_r$, a diffusion model is trained to transfer the input into partner feature using music $M$ and encode lead feature $F_l$ as the condition. This step ensures the output produces movements with the partner dancer characteristics.
\end{itemize}

\subsection{3D Pose Collection Stage}
Dance is composed of a series of basic dancing movement pieces. 
Thus, we follow the 3D Pose VQ-VAE in Bailando \cite{siyao2022bailando} to summarize the dance movement pieces in an unsupervised way. As shown in the left part of Figure \ref{overall}, a 3D Pose VQ-VAE \cite{yan2021videogpt} with Encoder and Decoder network is designed to collect numerous dance pieces. A 3D joint sequence $S$ is input into the Encoder with down sampling rate $d$ to obtain an encoded latent $F_{e} \in \mathbf{R}^{C \times N'}$, where $N'=N/d$. Each $d$ frame is regarded as a basic dance unit. Meanwhile, we randomly initialise two codebooks to collect dance movement pieces, one representing the upper body action $Z_u \in \mathbf{R}^{K \times C}$ and the other representing the lower body action $Z_d \in \mathbf{R}^{K \times C}$. Subsequently, we use $Z$ to denote both collectively, but in practice, the upper body and lower body vectors are separately computed with the codebook $Z_u$ and $Z_d$. By continuously updating with encoded latent $F_{e}$, these codebooks automatically store the dancing movement pieces as the training progresses. Then, the encoding latent $F_e$ is fed into a quantization operation to search for the nearest vector in the codebook, ensuring that the decoder can correctly recover the motion sequence from the code. The quantization operation can be formulated as:
\begin{equation}
  f_q^{n'}= \mathop{\arg\min} \limits_{\ z^{k} \in Z} \| f_e^{n'} - z^{k} \|,
\end{equation}
where $z^k$ means $K$-th vector of the codebooks and $n'$ is in the range of $N'$. 

Since modelling the upper and lower half bodies separately, we calculate the relative positions of each joint using the hip as the origin before encoding the input sequence $S$, to keep the coherence of the composed body. The original encoding latent $F_e$ is represented as a combination of series codes in the codebooks, which can be regarded as quantized latent $F_q$. Finally, the quantized latent $F_q$ is decoded by the Decoder to reconstruct the 3D joint sequence $\hat{S}$.


\textbf{Loss Function.} During the training stage, the annotations in the dataset are used as ground truth. The Encoder and Decoder are trained by minimizing the loss between the model output $\hat{S}$ and the corresponding ground truth $S$. The loss function followed by \cite{siyao2022bailando} is defined as:
\begin{equation}
  \mathcal{L}_{1} = \mathcal{L}(\hat{S},S) + \| \textbf{sg}(F_e) - F_q\| + \delta \|F_e-\textbf{sg}(F_q) \|,
\end{equation}
where $\textbf{sg}(\cdot)$ means the stop gradient. The loss function $\mathcal{L}(\hat{S},S)$ denotes the cumulative $\mathcal{L}_{1}$ loss between the predicted 3D joints and real joints, including their keypoints, velocities and accelerations. The accuracy of codebooks is constrained by the second loss $\| \textbf{sg}(F_e) - F_q\|$, while the third loss $\delta \|F_e-\textbf{sg}(F_q)\|$ is the commitment loss to optimize the Encoder with a weight $\delta$.
Overall, $\mathcal{L}_{1}$ loss is used to ensure that the codebook is updated correctly and can be decoded by a decoder.

\subsection{Dance Pre-generated Stage}
Now that any dance sequence can be represented by a series of quantized latent, the partner dancer generation task is then reframed to select proper latent from codebook $Z$ according to user-specified diversity, a given lead dancer and music. For any time $n'$, we estimate the latent with diffusion model and select the one with the nearest difference from $z^k \in Z$ as the generated pose latent. 

A Dance Pre-generated Diffusion model (DPGD) is introduced to pre-generate the diversity part as shown in Figure \ref{overall}. Given a dance sequence $S$ with the length of $N$, the upper and lower body poses are embedded to encoded features in the 3D Pose Collection stage and concatenated on the temporal dimension to obtain $F_q\in \mathbf{R}^{C \times (2\times N')}$. Then, aiming to control the diversity, we introduce a feature selection module to extract the preserved poses from concatenated feature $F_q$ by a mask. The reserved pose units are determined by the controllable parameters $\lambda$, which are determined via uniform selecting $2 \times N' \times \lambda$ units in the temporal dimension. For the rest of the unselected units, we set them as zeros and then get the selected feature $F_s$ with the same dimension of $F_q$.

With the selected latent $F_{s}$ as input, the proposed diffusion model is trying to fill this empty part as well as denoising. Diffusion model first add Gaussian noise $\epsilon \sim \mathcal{N}(0,\mathcal{I}) $ from $F_{s}^{0}$ to $F_{s}^{T}$ as:
\begin{equation}
 Q~ (F_{s}^t ~ | ~ F_{s}^{t-1}) = \mathcal{N}(F_{s}^t;~\sqrt{1-\beta_t}~F_{s}^t,~\beta_t \mathcal{I}),
\end{equation}
where $t$ means the time step of adding noise. $\beta$ is a set of constant hyper-parameters, each of which are ranged in $[0,1]$ and become larger as $t$ increases. Then, $F_{s}^{T}$ is inverse extrapolated to $F_{s}^{0}$ with a deep learning model as following:
\begin{equation}
  P~ (F_{s}^{t-1} ~ | ~F_{s}^{t}) = \mathcal{N}\big(F_{s}^{t-1};~\mu ~(F_{s}^{t},t),~\Sigma~(F_{s}^{t},t)\big),
\end{equation}
where $\mu$ and $\Sigma$ represent the mean and variance of the noised data distribution respectively. Here, we follow the training strategy in unCLIP \cite{ramesh2022hierarchical} to predict the initial input instead of directly modelling the noise $\epsilon$ as DDPM \cite{ho2020denoising}.

We conduct a U-Net structure \cite{ronneberger2015u} with the selected pose latent $F_{s}$ and time step $t$ as input for feature denoising. The noise time-step $t$ is projected into the latent space through a separate learned embedding layer and merged with the input $F_{s}$ feeding into U-Net. Since the transformer architecture enables learning the temporal information, we use two self-attention layers with five Resnet blocks as encoder and decoder. Every output $F_{r}$ in U-Net shared the same dimensions with the original pose latent.

\begin{figure*}[!t]
  \centering
  \setlength{\abovecaptionskip}{0.2cm}
  \includegraphics[width=0.9\linewidth, height=180pt]{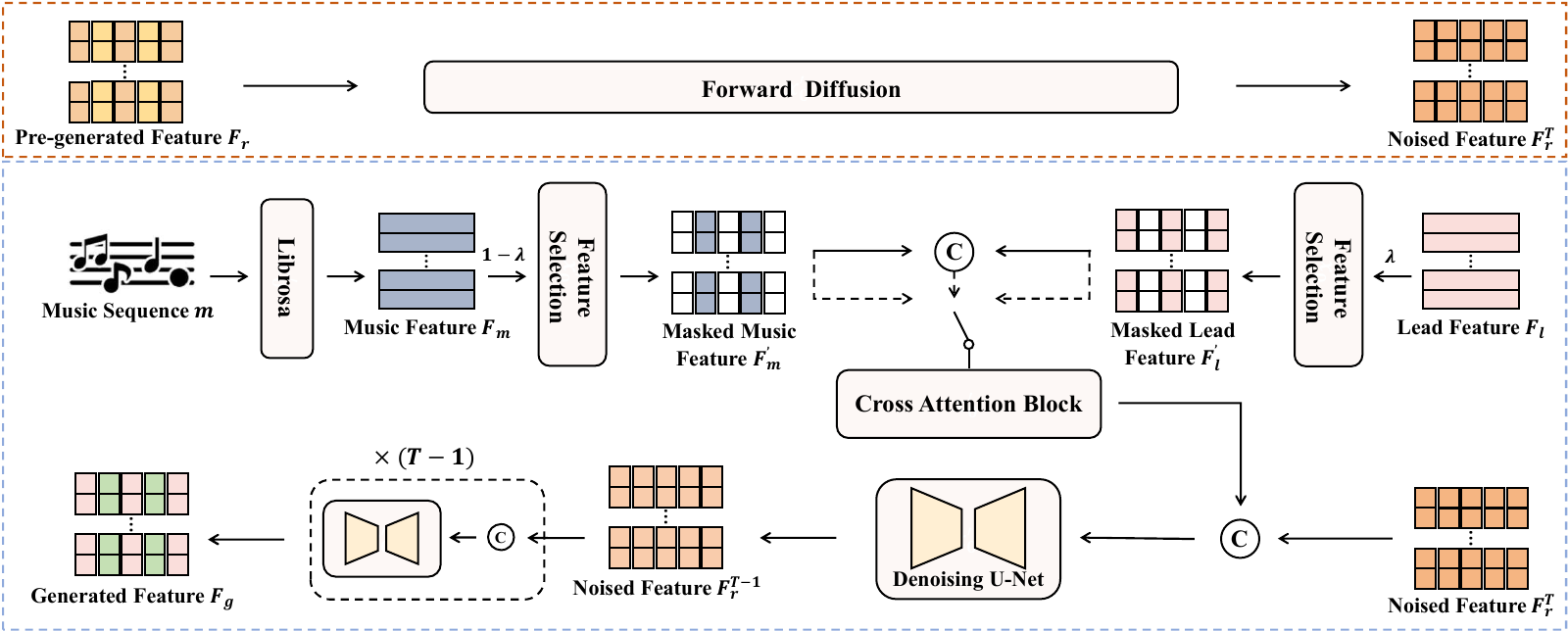}
  \caption{The architecture of Dance Motion Transfer stage. 
  We use classifier-free strategy \cite{ho2021classifier} to train the model, with 40\% probability of having one condition, 50\% probability of having two conditions, and the rest of having uncondition. Meanwhile, we utilize the given similarity to control which part of the condition should be updated when training.}
  \label{transfer}
\end{figure*}

\textbf{Loss Function.} The pre-generated diffusion model is optimized by minimizing the differences between output and lead dancer.
Concretely, the model is trained by the mean squared error (MSE) loss of the generated output $F_{r}$. Meanwhile, in order to control the similarity part (non-masked part) between the result and input as well as keep the diversity in the masked part, the loss function is defined as follows:
\begin{equation}
  \mathcal{L}_{2} = \|F_{r}^{\omega}-F_{q}^{\omega}\| + \gamma 
  \mathop{\min} \limits_{\ z^{k} \in Z} \|F_{r}^{1-\omega} - z^k \|.
\end{equation}
where $\gamma \in [0,1]$ is a trade-off, and $\omega$ means the timestamps of similarity part (also selected part).

The first part of the loss function is to count the MSE loss between the ground truth $F_{q}$ and the generated tensor $F_{r}$ in the selected timestamps. The second one is the \emph{codebook loss} to ensure the remaining motions are generated based on the previously collected. Without the \emph{codebook loss}, the unselected part may generate randomly smaller values, which is close to the same action in the codebook, resulting in a poor diversity of the final generation.

\textbf{Sampling.} According to DDPM \cite{ho2020denoising}, the sampling is done in an iterative manner, starting from $t=T$ and gradually sampling until back to the time step $t=0$. In practice, we train our U-Net model to learn the unconditional distributions and then gradually sample from noised feature $F^t_{p}$ in DDIM manner.

\subsection{Dance Motion Transfer Stage}
\label{sec3.3}
While the generated pose latent already meets similar requirements to the lead dancer after the previous stages, it is still lacking the relationship mapping between the lead dancer and partner dancer. Thus, we devise a Dance Motion Transfer Diffusion model (DMTD) to bring the pre-generated latent closer to the partner dancer sequences at the last stage. In more detail, we also apply a diffusion model with conditions to improve controllability. The forward Markov process remains the same in the second stage, with the difference that the pre-generated latent $F_{r}$ is used as a start. Meanwhile, DMTD models the distribution $p(F_{r}|F_{l},F_{m})$ as the reversed diffusion process that directly predicts the initial values.

Our model is illustrated in Figure \ref{transfer}, which uses two conditions as the generating guidance. For the condition part, the input music $m$ is firstly extracted to feature $F_m \in \mathbf{R}^{D\times N}$ by a public tool Librosa \cite{mcfee2015librosa}. Then, two conditions $F_l$ and $F_m$ are respectively fed into the feature selection module. The masked lead dancer feature $F'_l$ only guides the selected timestamp $\omega$ and the masked music feature $F'_m$ only guides the diverse generation of the rest $1-\omega$. Afterwards, two mask features are linearly projected into latent space and then fed into cross attention layer followed by concatenating with the input $F_r^{t}$. A U-Net, which shares the same structure with the DPGD model, is used to denoise the concatenated input at time step $t$. Besides, the noise time step $t$ is sinusoidal embedding into the model dimension, merging with conditional input at each Resnet Block, followed by FiLM \cite{perez2018film}. Moreover, we use the classifier-free strategy \cite{ho2021classifier} to train our model DMTD. The two conditions are not constantly given during the training process, and there is a probability that only one condition or zero condition is given. For conditions that are not specified, we do not activate their corresponding neurons in the training process.


\textbf{Loss Function.} To ensure that the U-Net captures different factors of dance sequences, similar to the Dance Pre-generated stage, the model output is divided into two components: the consistency component with the duration $\omega$ and the diversity component with the duration $1-\omega$. The model makes that the consistency component is close to the lead dancer feature, while the diversity component is close to the real partner dancer feature. Therefore, the loss function can be rewritten as:
\begin{equation}
  \mathcal{L}_{3} = \|F_{g}^{\omega}-F_{q}^{\omega}\| + \tau \|(F_{g}^{1-\omega}-F_{p}^{1-\omega})\|,
\end{equation}
where $F_{p}$ is the partner dancer feature corresponding to the lead dancer which is encoded in the 3D Pose Collection stage. $\tau$ is a trade-off weight.

\begin{figure*}[!t]
  \centering
  \setlength{\abovecaptionskip}{0.2cm}
  \includegraphics[width=0.9\linewidth]{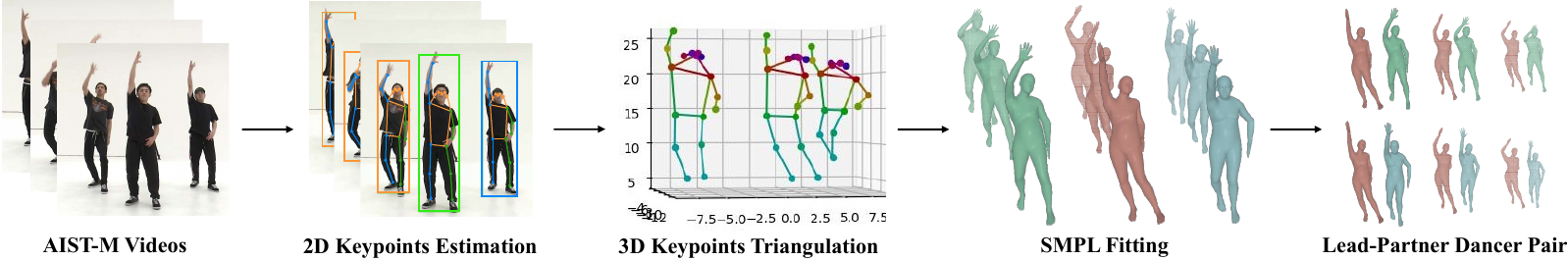}
  \caption{The pipeline of making AIST-M dataset. The different colors in SMPL fitting indicate different characters. 2D, 3D and SMPL skeleton is extracted from the video progressively. Finally, the dataset is divided into lead-partner dancer pairs.}
  \label{dataset}
  
\end{figure*}

\textbf{Sampling.} DMTD has the advantage of enabling controls for the similarity between dancers, which use the two conditions $F_l$ and $F_m$. Normally, the multiple constraints in the diffusion model are independent as done in \cite{liu2022compositional}. However, it does not meet our task requirements since there is a correlation between the given music and the lead dancer sequence. Hence, we argue that it is helpful to derive a generalized composing method. Similar as \cite{cho2023towards}, we formulate the sampling process as:
\begin{equation}
  p(F_{g}^t|F_l,F_m) \propto p(F_{g}^t) \big(p(F_l,F_m|F_{g}^t)^{\chi} \big(p(F_l|F_{g}^t)^{1-\varphi} p(F_m|F_{g}^t)^{\varphi} \big)^{1-\chi} \big)^{\alpha}, 
\end{equation}
where $\alpha \geq 0 $ controls the strength of conditions, while $\chi \in [0,1]$ is a trade-off weighting the independent effects and joint effects of the two conditions. $\varphi \in \mathbf{R}^{C \times (2\times N')}$ controls the weight for lead dancer and music information, which is a mask generated from the previously given similarity $\lambda$. The selected timestamps $\omega$ are set as 1. The rest of the timestamps $1-\omega$ are set as 0.
The parameter $\varphi$ ensures that when adjusting the trade-off $\chi$ of independent effects, the corresponding condition only controls the action at its relative timestamp, without affecting the other moments.

Based on this sampling process, the guidance gradient in terms of the denoising network $\epsilon_{\theta}$ (which may depend on zero, one or both conditions) is derived as follows:
\begin{equation}
\begin{split}
\nabla_{F_{g}^t}~logp~(F_{g}^t~|~F_1,F_m)&=\nabla_{F_{g}^t}~logp~(F_{g}^t,~F_1,~F_m) \\
  &=\epsilon_{\theta}(F_{g}^t,t)
   +\alpha \Big[ \chi \Big( \epsilon_{\theta}(F_{g}^t,t,F_l,F_m) - \epsilon_{\theta}(F_{g}^t,t) \Big) \\
  &+ (1-\chi) \Big(\varphi \big(~\epsilon_{\theta}(F_{g}^t,t,F_l) - \epsilon_{\theta}(F_{g}^t,t)\big) \\
  &+(1-\varphi)\big(\epsilon_{\theta}(F_{g}^t,t,F_m) - \epsilon_{\theta}(F_{g}^t,t)\big) \Big) \Big],
\end{split}
\end{equation}
where if $\chi = 0$, the process is simplified to generate sequences with two completely disjoint conditions. If $\chi = 1$, it means that the two conditions are highly related. 

\section{Dataset}


\subsection{Data Collection and Preprocessing}
We adopt the group dancing videos from the AIST dataset \cite{tsuchida2019aist}, each of which consists of three dancers without distinguishing roles. Table \ref{tab1} provides a comparison between our AIST-M dataset and similar open-source datasets. 


To annotate the dataset, we follow the pipeline used in the AIST++ dataset \cite{li2021ai}, depicted in Figure \ref{dataset}. For each perspective video, we utilize MMDetection \cite{chen2019mmdetection} and MMPose \cite{mmpose2020} to respectively obtain the tracking IDs and their COCO-format \cite{ruggero2017benchmarking} skeleton keypoints.
To improve the data quality, we apply a three-step pose optimization process. First, a sliding window approach is used to identify and remove outliers based on z-scores. Second, missing values are filled through interpolation using a polynomial function of degree 1. Finally, we utilize SmoothNet \cite{zeng2022smoothnet} to smooth the motion sequences. These preprocessing steps ensure that the 2D data is of high quality and suitable for further analysis.

\subsection{Lead-Partner dance Motion Annotation}
To construct full body motions, we first reorganize the tracking IDs from different perspectives into the correct person IDs. Then, the 2D multi-view keypoints are triangulated to obtain the 3D skeleton keypoints in COCO-format. To present the 3D human, we fit the 3D human body mesh vertices and joints using the SMPL model \cite{loper2015smpl},
Finally, we post-process and pairwise combine the annotated motion mesh in the same video, which removes the wrong cases and renames the remained combinations as the lead dancer and the partner dancer. 
\vspace{-1.0em}

\begin{table}[!h]
\setlength\tabcolsep{2.5pt}
\renewcommand\arraystretch{1}
\centering
\small
  \caption{Statistics of representative dance generation datasets. AIST-M focuses on classifying dancer roles and enables the investigation of partner dancer generation.}
  \vspace{-1.0em}
    \resizebox{\linewidth}{!}{
  \begin{tabular}{ccccccc}
    \toprule
    \multirow{2}{*}{Dataset} & \multirow{2}{*}{\begin{tabular}[c]{@{}c@{}}Time\\ (h)\end{tabular}} & \multirow{2}{*}{\begin{tabular}[c]{@{}c@{}}Dance\\ Styles\end{tabular}}
    & \multirow{2}{*}{\begin{tabular}[c]{@{}c@{}}2D\\ Joint\end{tabular}}
    & \multirow{2}{*}{\begin{tabular}[c]{@{}c@{}}3D\\ Joint\end{tabular}}
    & \multirow{2}{*}{\begin{tabular}[c]{@{}c@{}}Group\\ Dance\end{tabular}}
    & \multirow{2}{*}{\begin{tabular}[c]{@{}c@{}}Lead-Partner\\ Dancer Pair\end{tabular}}  \\    \\
    \midrule
    Dance with Melody \cite{tang2018dance} & 1.57 & 4 & \ding{55} & \ding{51} & \ding{55} & \ding{55} \\
    DanceNet \cite{zhuang2022music2dance} & 0.96  & 2 & \ding{55} & \ding{51} & \ding{55} & \ding{55}\\
    Dancing2Music \cite{lee2019dancing}& 71 & 3 &  \ding{51} & \ding{55} & \ding{55} & \ding{55}\\
    Dance Revolution \cite{huang2020dance} & 12 & 3 &  \ding{51} & \ding{55} &  \ding{55} & \ding{55}\\
    AIST++ \cite{li2021ai}& 5.2 & 10 &  \ding{51} &  \ding{51} &  \ding{55} & \ding{55}\\
  \textbf{AIST-M (ours)} & 1.02 & 10 &  \ding{51} &  \ding{51} &  \ding{51} &  \ding{51} \\
  \toprule
\end{tabular}}
\label{tab1}
\vspace{-2.0em}
\end{table}

\section{Experiments}

\subsection{Implementation Details}

\textbf{Dataset.} We perform the training and evaluation on the proposed dataset AIST-M, in which we totally collect 340 lead-partner dancer pairs and randomly select 40 pieces of different styles for test and the rest for training. Besides, the music is extracted by the public audio processing toolbox Librosa \cite{mcfee2015librosa} to obtain spectral, onset, rhythmic and beat feature, including Mel Frequency Cepstral Coefficients, Constant-Q Chromagram, Onset Strength, Onset, Tempogram and Beat Tracker. The extracted music features have a total of 419 dimensions.
     
     
     


\begin{figure*}[!t]
  \centering
  \setlength{\abovecaptionskip}{0.2cm}
  \includegraphics[width=0.9\linewidth,height=220pt]{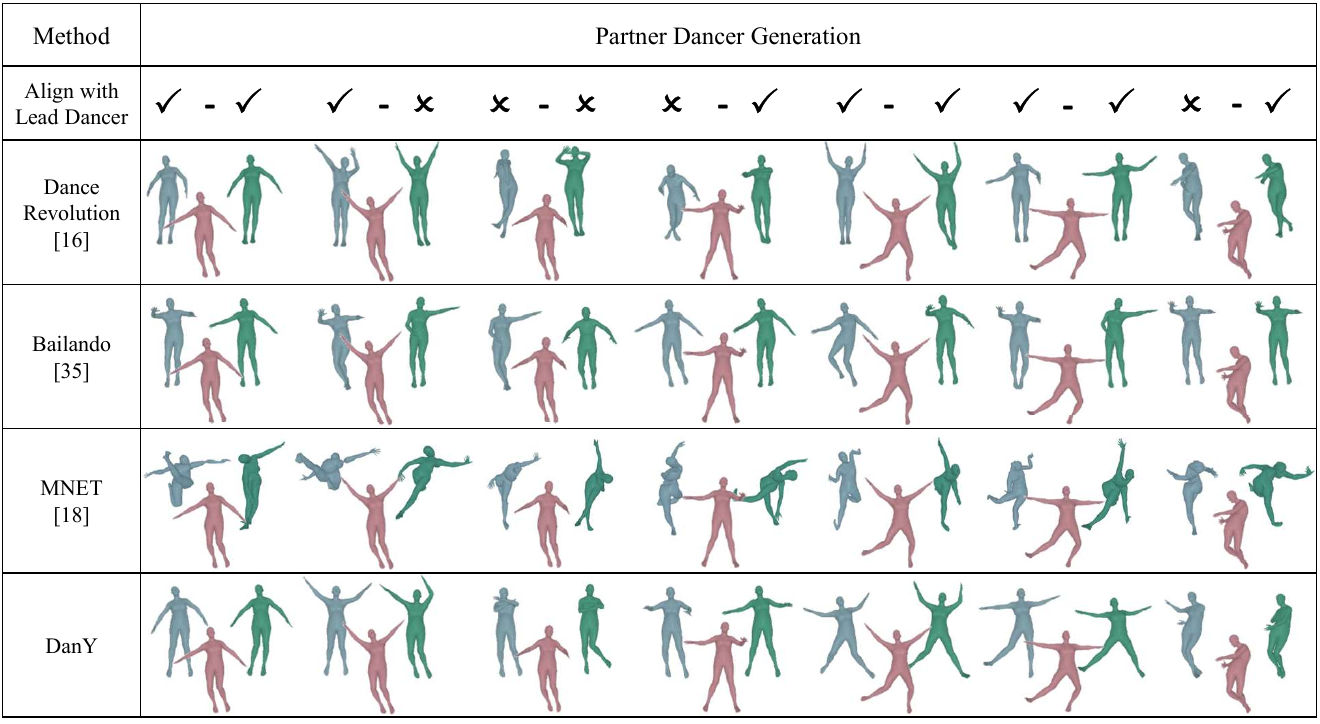}
  \caption{Visualization of partner dancer generation by our DanY and baseline methods. The pink person represents the lead dancer sequences, while the blue and green one represents the result with similarity parameters $\lambda$ 0.2 and 0.5 respectively. The second row indicates whether this frame should be aligned with the lead dancer, using check mark to represent align and cross mark to represent no need for consistency.}
  \label{result}
\end{figure*}

\textbf{3D Pose Collection Stage.} The initial dance motions are cropped to $N=256$ in VQ-VAE training. The dimension of both upper and lower pose codebook $K$ is set to 512 as well as the dimension of the encoded latent feature $C$. Meanwhile, the temporal downsampling rate $d$ of encoders is set to 8 and the batch size is 32. To preserve the motion information of single dancer, we fine-tune the \cite{siyao2022bailando} pre-train model on the AIST++ dataset. Adam optimizer \cite{kingma2014adam} is adopted with the learning rate $3 \times 10^5$, $\beta_1 = 0.9$ and $\beta_2 = 0.99$ to train multiple person pose VQ-VAE for 500 epochs. The trade-off $\delta$ is set to 0.1 as the original configuration. 

\textbf{Dance Pre-generated Stage.} The quantized pose code is generated by VQ-VAE with the temporal dimension $N'=32$. The batch size is set to 64 and $\gamma$ is 0.1 in training. The DPGD model is optimized using stochastic gradient descent (SGD) \cite{ruder2016overview} optimizer with an initial learning rate of 0.005 and momentum of 0.9 for 200 epochs, and the learning rate decays 0.1 every 50 epochs. While sampling, we use DDIM to directly sample the masked pose code with $\eta= 0 $, and iterative denoise 10 steps.

\textbf{Dance Motion Transfer Stage} The model is adopted classifier-free guidance \cite{ho2021classifier} in this stage. The training time for the diffusion model without conditions, with two conditions and with three conditions are 20\%, 20\% and 50\% of the overall epochs. We train the model with a batch size of 64 and a latent dimension of 512 for 200 epochs. The optimizer settings are the same as the DPGD model, and the trade-off $\tau$ is set to 2. While sampling, we evaluate our models with guidance scale $\alpha=9$, $\chi=0.9$ as the baseline default, and use reverse DDIM sampling with step $t=50$.

\subsection{Evaluation Metrics}
Frechet Inception Distance (FID) \cite{heusel2017gans} is an algorithm commonly used to judge the similarity between two poses in motion generation, while it is not suitable for evaluating the similarity in partner dancer generation task.
Therefore, we utilize a novel metric Merge Frechet Inception Distance (MFID), which is an extension of FID, to synthetically evaluate the similarity to the lead and partner dancers. MFID both considers the lead dancer and partner dancer by separately calculating similar part sequences and diverse part sequences.
\begin{equation}
 MFID = \frac{1}{2} \big( \{FID(\hat{S}^i,S_l^i) | i\in N_l \}+ \{ FID(\hat{S}^j,S_p^j)| j \in N_p \} ),
\end{equation}
where $N_l$ and $N_p$ represent the frame index that should be consistent with the lead dancer and the real partner dancer, respectively. $\hat{S}$ is the output in our framework.

The diversity of motion sequence is also worth considering. Generally, generation diversity (GenDiv) \cite{lee2019dancing,huang2020dance} is used to evaluate the variety between poses. Regarding the alignment between music and generated motions, we use the Beat Align Score in \cite{siyao2022bailando}, which is the average temporal distance between each music beat and its closest dance beat. Therefore, these two metrics are also used to evaluate the generation result.


\subsection{Experimental Results}
We compare our proposed model to several individual dance generation methods including Dance Revolution \cite{huang2020dance}, Bailando \cite{siyao2022bailando} and MNET \cite{kim2022brand}. Specifically, we follow the baseline methods as: (1) Dance Revolution framework for multiple dancers with 3D pose VQ-VAE; (2) Bailando framework for multiple dancers without reinforcement learning; (3) MNET framework for multiple dancers; (4) our entire framework. For each method, we use the same set of codebooks and set the similarity parameters $\lambda$ as 0.5.

\textbf{Qualitative Evaluation.}
Figure \ref{result} shows the qualitative results. 
This indicates that the previous single dancer methods failed to achieve controllable diversity, since the occurrence of consistent motion differ from the specified timestamps, and it can even lead to distorted generated motions (such as the results in MNET). In contrast, the last row of Figure \ref{result} shows the effectiveness of our DanY framework. Most importantly, even in the aligned frames, the results we generate are highly similar but not identical to the lead dancer sequence, which indicate that our proposed framework can effectively generate diverse motions.

\textbf{Quantitative Evaluation.}
Table \ref{com_other} shows the quantitative results. According to the comparison, our proposed model outperforms all the other baselines on FID evaluations. Furthermore, our framework achieves a higher GenDiv, indicating fewer repetitive actions in the generated sequence and a lower probability of stiffness. This enables the efficient generation of continuous and diverse dancer sequences in our DanY. Besides, the value of Beat Align Score in our DanY indicates that our framework is better able to synchronize the dancer movements with the rhythm of the music, resulting in more expressive and natural partner dancers that follow the tempo and timing of the music more closely.

\begin{table}[!h]
 \setlength\tabcolsep{2.5pt}
 
\renewcommand\arraystretch{1}
\centering
\setlength{\abovecaptionskip}{0.2cm}
\small
  \caption{Comparison of our framework and individual dance models. The similarity $\lambda$ is set to 0.5. The MFID reflects generating results in our framework has both lead and partner dancer movements. The GenDiv indicates the richness of our generated poses, as well as aligns with the music.}
  \label{tab3}
  \resizebox{\linewidth}{!}{
  \begin{tabular}{ccccc}
    \toprule
     Methods & Venue & MFID $\downarrow$  & GenDiv $\uparrow$ & Beat Align Score $\uparrow$  \\
    \midrule
    Dance Revolution \cite{huang2020dance} & ICLR'21 & 67.73 & 11.27 & 0.234  \\
    Bailando \cite{siyao2022bailando} & CVPR'22 & 236.84 & \textbf{11.40} & 0.238 \\ 
    MNET \cite{kim2022brand} & CVPR'22 & 6045.38  & 7.99 & 0.230  \\
    Ours  & - & \textbf{40.25} & \textbf{11.40}  &\textbf{0.240}\\
     \toprule
\end{tabular}
}
\label{com_other}
\end{table}

\subsection{Ablation Study}

\textbf{The Similarity Parameters Analysis.}
Table \ref{com_sim} provides a comparison between different settings of similarity, including the evaluation of the lead dancer sequence as the lower bound. The results show that our generated sequences have higher pose diversity and better matching with the music than the given dance sequence in all similarity settings. Besides, as $\lambda$ increases, the number of non-similar postures decreases, resulting in a decrease in the MFID, which implies an increase in the similarity with the original lead dancer's movements. As $\lambda$ decreases, the GenDiv is increasing, which means the generated results have a high diversity compared to the original partner dancer sequence for the timestamps inconsistent with the lead dancer. Moreover, the Beat Align Score is influenced by the similarity parameter, as higher or lower values of similarity result in less alignment with the music. Our results demonstrate that the appropriate similarity parameter can balance the consistency of the lead dancer and the matching with the music. Furthermore, Figure \ref{sim} shows the results under different similarity settings at the same timestamp, which corroborates that the higher the similarity, the higher the probability of generating a result with a similar pose to the lead dancer.

\begin{table}[!h]
\setlength\tabcolsep{5pt}
\setlength{\abovecaptionskip}{0.2cm}
\renewcommand\arraystretch{1}
\centering
\small
  \caption{Ablation study on the similarity parameter $\lambda$. The lower bound is the evaluation result of lead dancer sequences. When the value of similarity increases, the generated sequences are diverse. The degree of music-action alignment peaks at around the median of similarity.}
  \begin{tabular}{cccc}
    \toprule
    The value of similarity  & MFID $\downarrow$ & GenDiv $\uparrow$ & Beat Align Score $\uparrow$ \\
\midrule
    0 & 143.11 & \textbf{11.45} & 0.223\\
    0.2 & 89.93 & 11.39 &0.231  \\
    0.6 &44.71 & 11.21 & \textbf{0.257}  \\
    0.8 & 8.64 & 11.27&0.214    \\
    1  & \textbf{3.75} & 11.20& 0.217  \\
    \midrule
    Lower Bound & - & 11.10 & 0.205\\
     \toprule
\end{tabular}
\label{com_sim}
\end{table}

\begin{figure}[!t]
  \centering
  \setlength{\abovecaptionskip}{0.1cm}
  \includegraphics[width=0.9\linewidth]{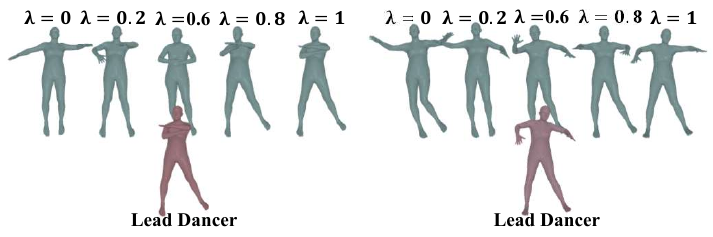}
  \caption{Visualization of partner dancer generation by our DanY with different similarities. The red person is the given lead dancer sequence and the rest are the generated results under different degrees of similarity $\lambda$. }
  \label{sim}
\end{figure}

\textbf{Three-Stage framework DanY.}
In Table \ref{com_con}, we evaluate the contribution of different parts in our DanY with the following methods: (1) using the first two-stage; (2) using the first and third stages; (3) using all three stages. The similarity is set to 0.6 for all experiments. 
 We first evaluate the result generated in Dance Pre-generated stage, which shows a higher MFID, indicating that the stage is less similar to the partner dancers. When the framework is set without Dance Pre-generated stage, although it achieves the highest Beat Align Score, its MFID vaule is slightly difference with the results directly obtain in Dance Pre-generated stage. This indicates that the results can only guarantee the similarity of the lead dancer and cannot closely replicate the real partner dancer movements. Overall, the ablation study confirms the importance of each stage in our proposed framework and demonstrates its effectiveness in generating diverse and consistent partner dancers.

\begin{table}
\setlength\tabcolsep{5pt}
\renewcommand\arraystretch{1}
\setlength{\abovecaptionskip}{0.2cm}
\centering
\small
  \caption{Ablation study on our framework. Experiments are conducted on Dance Pre-generated stage and Dance Motion Transfer stage, respectively. The similarity $\lambda$ is set to 0.6.}
  \resizebox{\linewidth}{!}{
  \begin{tabular}{ccccc}
    \toprule
     & Methods & MFID$\downarrow$  & GenDiv$\uparrow$ & Beat Align Score$\uparrow$\\
    \midrule
    (1) &Ours \textit{w/o} DMTD Model   & 44.80 &\textbf{11.21} & 0.259   \\
     (2) & Ours \textit{w/o} DPGD Model &44.79 &\textbf{11.21} & \textbf{0.260}  \\
    
    (3) & Ours & \textbf{44.71} & \textbf{11.21} & 0.257 \\
    \toprule
\end{tabular}
}
\label{com_con}
\vspace{-1.5em}
\end{table}

\section{Conclusion}
In this paper, we proposed a novel task of partner dancer synthesis, which generates partner dancers with controllable diversity while maintaining temporal coordination with the lead dancer. To address the lack of suitable datasets, we construct the AIST-M dataset, which includes lead-partner dancer pairs in various formats. Moreover, we devise a three-stage framework to automatically generate partner dancers with given similarity constraints from input lead dancer sequences. Additionally, the sampling method is improved in the diffusion model to ensure the efficient mixing of multiple conditions. Our extensive experimentation on the AIST-M dataset provides strong evidence for the potential of our DanY for generating expressive and realistic partner dancers with controllable diversity.

\begin{acks}

The work is partially supported by the Young Scientists Fund of the National Natural Science Foundation of China under grant No. 62106154, by the Natural Science Foundation of Guangdong Provin-
ce, China (General Program) under grant No.2022A1515011524, and by Shenzhen Science and Technology Program 
JCYJ2022081810300-1002 and by Shenzhen Science and Technology Program ZDSYS202-11021111415025.
\end{acks}

\clearpage


\bibliographystyle{ACM-Reference-Format}
\balance
\bibliography{sample-base}


\clearpage

\appendix

\section{AIST-M Dataset Details}
\subsection{AIST-M Dataset Statistics} 
Our dataset contains in total of 1.02 hour of different 3D dance motions accompanied by music which are then reconstituted into 340 lead-partner dancer pairs. The dataset covers 10 dance genres and 60 pieces of music (only 51 pieces are used in lead-partner dancer pairs). Among the motion sequences for each genre, the duration of each motion sequence is ranging from 29 seconds to 48 seconds. 

\begin{figure}[!h]
  \centering
    \setlength{\abovecaptionskip}{0.1cm}
  \includegraphics[width=0.65\linewidth]{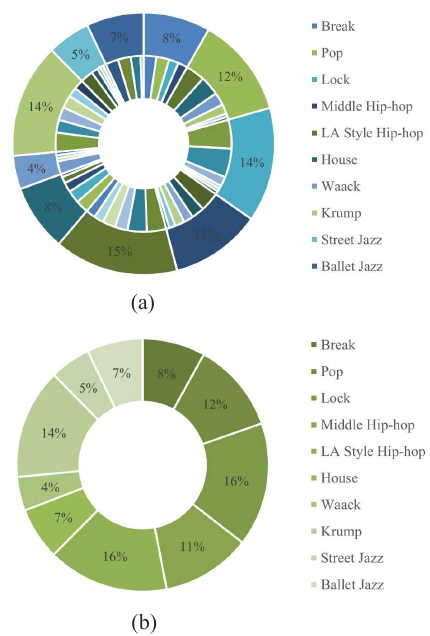}
  \caption{Distribution (\%) of music pieces (a) and Lead-Partner dancer sequences (b) in our AIST-M dataset. The outermost circle of (a) represents the number of motion sequences in each genre, and the inner circle indicates the percentage of the whole for each different piece of music. (b) illustrates the distribution of the duration of the Lead-Partner dancer pair for different dance genres.}
  \label{bing}
  \vspace{-0.2cm}
\end{figure}

Figure \ref{bing} show the detailed distribution of music pieces and motion sequences for each genre in our AIST-M dataset. As illustrated in Figure \ref{bing}-(a), the outermost circle is the number of data in each genre of lead-partner dancer pairs, and the inner circle indicates the percentage of the whole for each different piece of music. Each genre contains 4 to 6 different music pieces, of which LA Style Hip-hop accounts for up to 15\% and Waack accounts for a minimum of 4\%. Figure \ref{bing}-(b) shows the distribution of the duration of the Lead-Partner dancer pair for different dance genres. Lock, LA Style Hip-hop, and Krump are dominant dance genres.

\subsection{AIST-M Motion Diversity Visualization} 
\begin{figure}[!h]
  \centering
  \setlength{\abovecaptionskip}{0.1cm}
  \includegraphics[width=1\linewidth]{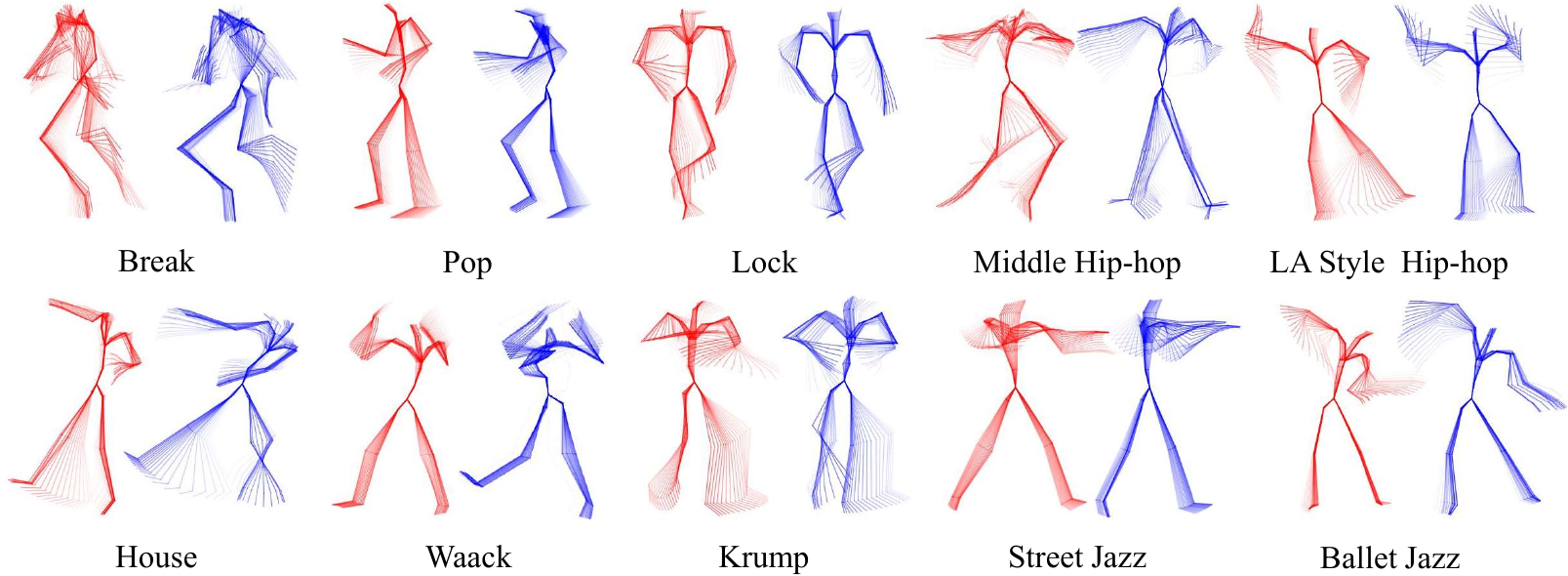}
  \caption{Visualization of 10 genres in AIST-M dataset. The red person represents the lead dancer and the blue person represents the partner dancer.}
  \label{genre}
  \vspace{-0.5cm}
\end{figure}

In Figure \ref{genre}, we show an example action from 10 genres in our AIST-M dataset. For each action, we have two people, the lead dancer (red one) and the partner dancer (blue one). It can be seen in the performance of the same action, the lead dancer and the partner dancer's posture may be the same, and there may also be a mirror posture for better cooperation, or similar posture due to the difference in speed.

\subsection{AIST-M Motion Accuracy of Annotation} 
\begin{figure}[!h]
  \centering
  \setlength{\abovecaptionskip}{0.0cm}
  \includegraphics[width=0.65\linewidth]{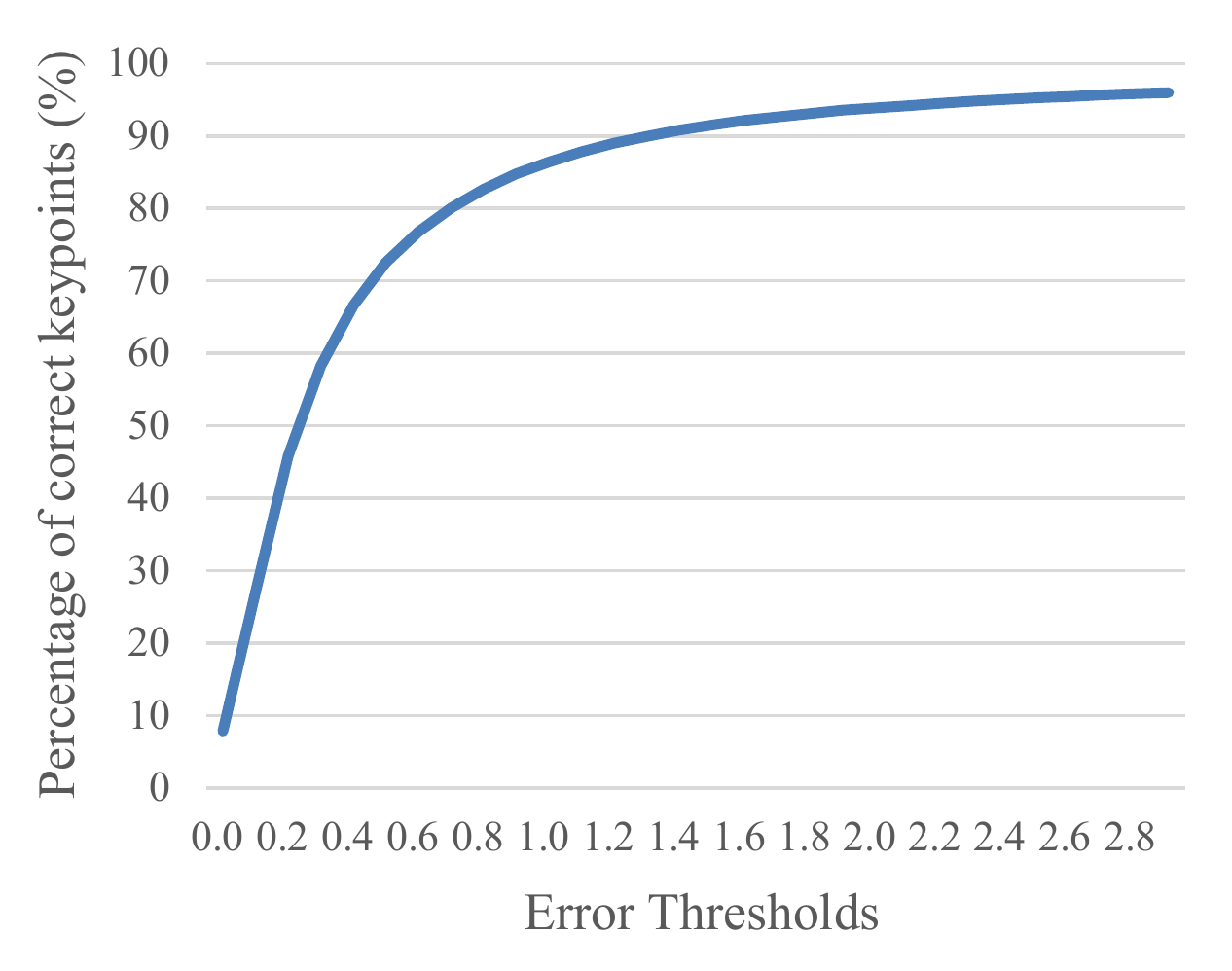}
  \caption{PCKh Metric on AIST-M. We analyze the PCK metric between re-projected 2D keypoints and detected 2D keypoints on AIST-M. We use body length as a scaling factor for evaluation. }
  \label{pck}
  \vspace{-0.5cm}
\end{figure}

Figure \ref{pck} shows the Percentage of Correct Keypoints (PCK) \cite{modec13} metric on AIST-M to show the annotation validity in our AIST-M dataset. We re-project the 3D keypoints to 2D and compare them with the original detected 2D keypoints. Averaged PCK at the threshold of 2.4 is 95.0 \% on all joints shows that our reconstructed 3D keypoints are consistent with the predicted 2D keypoints, where the difference is due to the presence of outlier points and different distances of people from the camera. 

\subsection{Extracted Music Feature} 
Table \ref{tab2} details the extracted features of music that are fed into our model. For each genre, the pieces of music are ranging from 29 seconds to 54 seconds long, and from 80 BPM to 130 BPM (except for the House genre which is 110 BPM to 135 BPM). According to their Spectral, Onset, Rhythmic and Beat features, we extract Mel Frequency Cepstral Coefficients, Constant-Q Chromagram, Onset Strength, Onset, Tempogram and Beat Tracker with a total of 419 dimensions of feature information.

\begin{table}[!h]
    \setlength\tabcolsep{5pt}
    \setlength{\abovecaptionskip}{0.2cm}
    \renewcommand\arraystretch{1}
    \centering
    \small
    \caption{The music feature extraction by Librosa \cite{mcfee2015librosa}. According to Spectral, Onset, Rhythmic and Beat, we extracted a total of 419-dimensional features.}
    \resizebox{\linewidth}{!}{
      \begin{tabular}{ccc}
        \toprule
        Feature Categories & Feature Name & Dimension \\
        \midrule
        \multirow{2}{*}{Spectral Feature} & Mel Frequency Cepstral Coefficients & 20 \\
        ~& Constant-Q Chromagram  & 12  \\
         
        \multirow{2}{*}{Onset Feature}  & Onset Strength & 1  \\
        ~ & Onset & 1  \\
         
        Rhythmic Feature & Tempogram & 384   \\
         
        Beat Feature & Beat Tracker & 1   \\
        
      \midrule
      Total & \ & 419  \\
      \toprule
    \end{tabular}}
    \label{tab2}
    \vspace{-0.2cm}
\end{table}

\section{Framework Details}
\subsection{The Visualization of Collected Code}
\begin{figure}[!h]
      \centering 
      \setlength{\abovecaptionskip}{0.2cm}
      \includegraphics[width=0.8\linewidth,height=110pt]{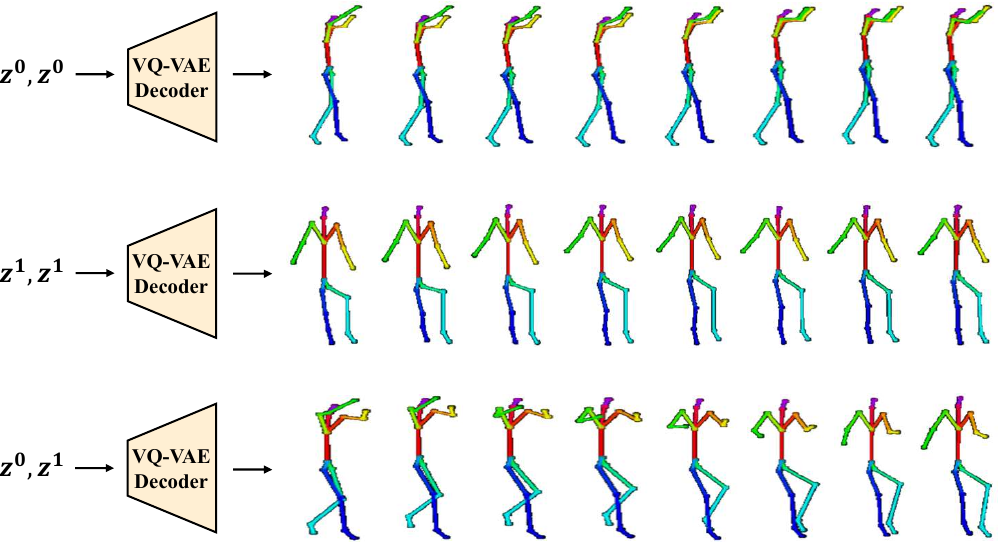}
      \caption{Visualization of collected code in the 3D Pose Collection stage. $z^0$ and $z^1$ are two different codes. The sequence of a single code is decoded to a relatively static pose, while the sequence of two various codes is decoded to smooth the transition between two poses.}
      \label{code}
      \vspace{-0.2cm}
\end{figure}

In the 3D Pose Collection stage, we propose to summarize meaningful dancing units into the codebook via a VQ-VAE. To demonstrate that our summarized codes are able to collect different dance units and to produce smooth movements, we visualize the latent in the codebook as shown in Figure \ref{code}. $z^0$ is set to $[22,22]$ and $z^1$ is set to $[124,124]$, where the numbers represent upper and lower body separately. The first two rows indicate the results of using the same set of codes, and it can be noticed that the actions of the two results are almost constant. The third row indicates the result of using two different sets of codes, which shows the resulting action gradually changing from one action to another. This image explains well that, for an arbitrary combination of learned codes in the codebook, the decoders can synthesize fluent movement based on the represented dance codes.

\subsection{The Discussion of Hyper-parameter in Dance Motion Transfer Stage} 
\begin{figure}[!hb]
      \centering 
      \setlength{\abovecaptionskip}{0.2cm}
      \includegraphics[width=0.9\linewidth]{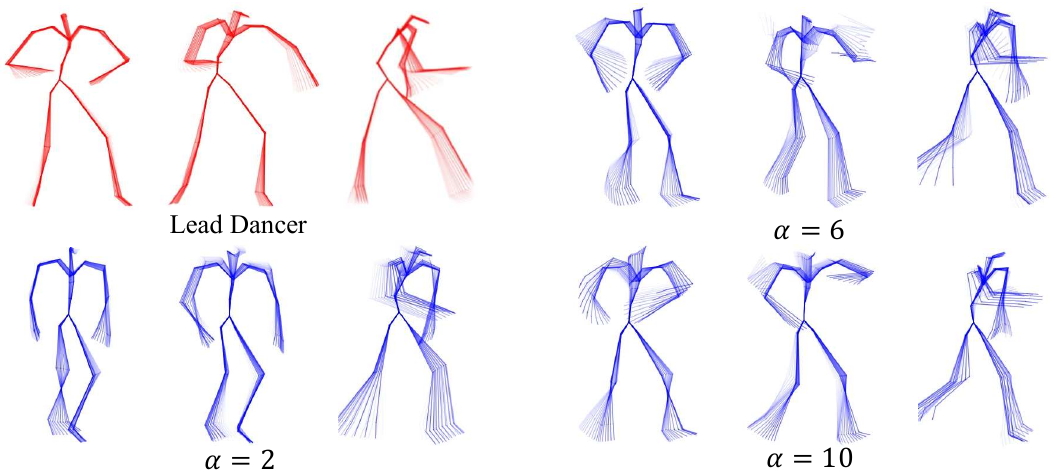}
      \caption{Visualization of different value of $\alpha$ in the Dance Motion Transfer stage. The red person represents the lead dancer, while the blue person represents the generated dancer with a different setting. The $\chi$ is set to 0.9 and similarity $\lambda$ is 0.2.}
      \label{gs}
\end{figure}

\begin{figure}[!hb]
      \centering 
      \setlength{\abovecaptionskip}{0.2cm}
      \includegraphics[width=0.9\linewidth]{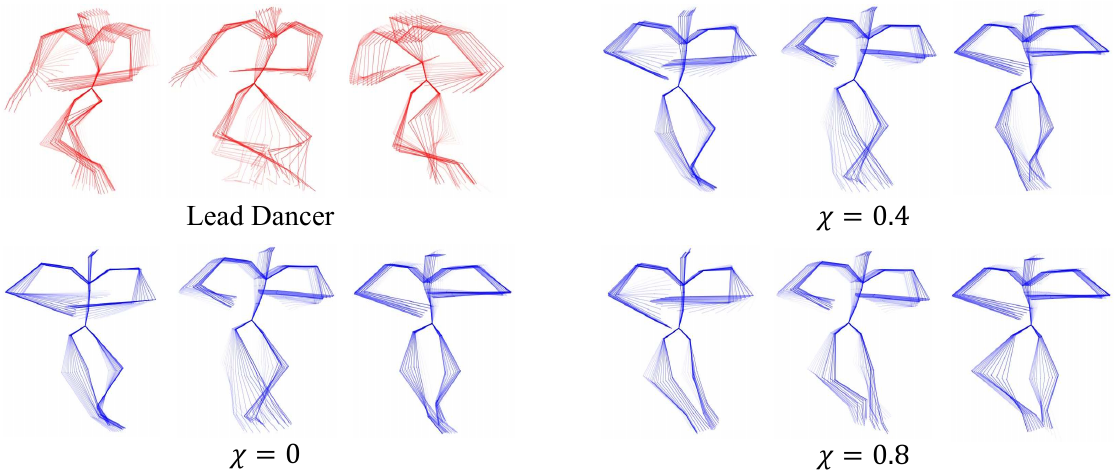}
      \caption{Visualization of different value of $\chi$ in the Dance Motion Transfer stage. The red person represents the lead dancer, while the blue person represents the generated dancer with a different setting. The guidance scale $\alpha$ is set to 9 and similarity $\lambda$ is 0.5.}
      \label{chi}
\end{figure}

In the Dance Motion Transfer stage, two hyper-parameters, $\alpha$ and $\chi$, are proposed to separately control the impact degree of conditions and the correlation between conditions. Figure \ref{gs} presents the result with different guidance scales $\alpha$. With $\alpha$ increasing, the posture similarity between the red dancers and the blue dancers increases, but this similarity decreases as $\alpha$ reach overly large, resulting in diverse posture such as mirroring. Figure \ref{chi} shows a visualization of different values of $\chi$, which indicates that the generated results become more similar to the posture of the lead dancer as $\chi$ increases.  

\subsection{The Discussion of Different Dance Genres} 
Different dance genres exhibit different challenges in the partner dancer generation. As shown in Table \ref{tabstyle}, LA Style Hip-pop and Ballet Jazz are easier for the model to generate, while Break and Middle Hip-pop contain highly skilled movements that are challenging to capture and represent accurately. In addition, our framework is capable of ensuring diversity and alignment with the music when generating movements in various genres. More visualization results show in the supplementary video.

\begin{table}[!b]
\setlength\tabcolsep{5pt}
\setlength{\abovecaptionskip}{0.2cm}
\renewcommand\arraystretch{1}
\centering
\small
\caption{The results of different dance genres. These results are obtained with similarity $\lambda$ 0.5 by a same model.}
\resizebox{\linewidth}{!}{
  \begin{tabular}{cccc}
    \toprule
    Genres & MFID $\downarrow$  & GenDiv $\uparrow$ & Beat Align Score $\uparrow$ \\
    \toprule
    Break& 96.76& 11.27& 0.24\\
    Pop& 69.90& 11.21& 0.24 \\
    Lock& 56.92& 11.29& 0.25 \\
    Middle Hip-pop& 93.46& 11.45& 0.23\\
    LA Style Hip-pop& 40.61& 11.25& 0.26 \\
    House& 61.67& 11.32& 0.20\\
    Waack& 71.40& 11.29& 0.24\\
    Krump& 40.95& 11.24& 0.24\\
    Street Jazz& 48.15& 11.20& 0.23\\
    Ballet Jazz& 41.12& 11.30& 0.23\\
\bottomrule
\end{tabular}}
\label{tabstyle}
\vspace{-0.5cm}
\end{table}

\end{document}